\documentstyle[aps]{revtex}
\title{\bf  All Optical Cellular Quantum Computer having Ancilla Bits for Operations in Each Cell}
\draft
\author{Toshio Ohshima}
\address{Fujitsu Laboratories Ltd.,
10-1 Morinosato-Wakamiya, Atsugi 243-0197, Japan}
\date{Received: \today}
\begin{document}
\maketitle
\begin{abstract}
A quantum cellular network with a qubit and ancilla bits in each cell is proposed. The whole circuit works only with the help of external optical pulse sequences. In the operation, some of the ancilla bits are activated, and autonomous single- and two-qubit operations are made. In the sleep mode of a cell, the decoherence of the qubit is negligibly small. Since only two cells at most are active at once, the coherence can be maintained for a sufficiently long time for practical purposes. A device structure using a quantum dot array with possible operation and measurement schemes is also proposed. 
\end{abstract}

Since the formulation of the quantum computer and quantum circuit \cite{Deutsch}, theoretical studies on quantum computers have clarified their efficiency in highly complex computational tasks; factorization and search, are two examples \cite{DJ,Shor,Grover}. Some physical implementations have been proposed. Moreover, their operations on a few qubits have been demonstrated experimentally\cite{Cirac,Cory1,Gershenfeld1,Gershenfeld2,Cory2,Jones,Reck,Cerf,Takeuchi}. Although these studies are impressive, these systems (e.g.,liquid NMR, ion trap, linear optics) are not promising since they are limited in their possible integration: 10 qubits or thereabouts. 

Thus, a quantum computer based on conventional semiconductor technology  has been expected. Kane proposed a silicon-based qubit using the nuclear spin of a phosphorus donor \cite{Kane}. Although this idea has another advantage of the good isolation of nuclear spins from the environment, it requires precise position control of individual impurities. On the other hand, qubits based on the electrons in semiconductor nanostructures are more realistic since, nowadays, quasi-zero dimensional quantum dots can be fabricated with high feasibility using various techniques. Furthermore, the control of the motion of individual electrons in such structures was also proved to be possible using single electron effects.

In 1995, Barenco et al. proposed a quantum dot-based implementation \cite{Barenco}. They utilized the quantized energy levels in a quantum dot as two states of a qubit. The coulomb interaction between dipole moments of two adjacent dots induced by an externally applied electric field was used for the controlled NOT operation. The observation of discrete levels and the control of electron number in quantum dots have been performed experimentally \cite{Tarucha,Oosterkamp}. It should become possible to control coherently the quantum states from outside. However, in general, the decoherence of electron wavefunctions in excited levels is fast since they are strongly coupled to electromagnetic and acoustic environments. The quantum error correction and fault-tolerance theories have shown that some amount of decoherence can be overcome \cite{Shor:error,Steane1,Calderbank,Steane2,Laflamme,Bennett,DiVincenzo,Steane3,Preskill}. However, a sufficiently large time ratio, $R_t$ of coherence time to gating time, is required in order for the correction procedures to be effective. The simple quantum dot scheme does not fulfill this criterion. 

In 1998, Loss and DiVincenzo proposed to use the spin of an electron confined in a quantum dot \cite{Loss,Burkard}. The spin degree of freedom is expected to be well isolated from photon and phonon environments. Furthermore, the two-qubit operation is performed by simple spin swap dynamics instead of the usual controlled-NOT. This operation is virtually a two-electron quantum beat, which is switched on by lowering the barrier between two quantum dots to enable interdot tunneling. The two spins of quantum dots are exchanged continuously in time. This scheme, however, has difficulty in the modulation of the barrier. Magnetic gating has a disadvantage in the operation speed, which reduces the time ratio eventually. Gating by nanoscale electrodes requires an extremely advanced fabrication technology. 

In this paper, we propose a new, all-optical implementation scheme for a quantum dot-based quantum circuit that has the cellular structure of a qubit and ancilla bits in each cell. In our scheme, single- and two-bit operations are made if and only if specific ancilla bits are excited (the active state of a cell). In the ground state of a cell, no operation is made (the sleeping state of a cell). Therefore, in sleeping cells, the coherence is maintained very faithfully. Since the duty ratio of active time to total time scales as $N^{-1}$, the effective time ratio becomes N times larger.

In our cell, the Hilbert space is spanned by the kets:
\begin{eqnarray}
|w,z_1,z_2,\ldots,z_m\rangle=|w\rangle\otimes|z_1\rangle\otimes|z_2\rangle\otimes\cdots\otimes|z_m\rangle,
\end{eqnarray}
where $|w\rangle$ denotes the qubit state and $|z_k\rangle, (k=1,\cdots,m)$ are the states of ancilla bits. Thus, our cell is a $2^{m+1}$-level system. However, by adopting a specific physical design and operation scheme, which will be described later, we can ignore most parts of the Hilbert space. As a result, we may consider only 
\begin{eqnarray}\label{contract}
\begin{array}{lcl}
|w,0\rangle&=&|w,0,0,\ldots,0\rangle\\
|w,1\rangle&=&|w,1,0,\ldots,0\rangle\\
|w,2\rangle&=&|w,0,1,\ldots,0\rangle\\
&\vdots&\\
|w,m\rangle&=&|w,0,0,\ldots,1\rangle,

\end{array}
\end{eqnarray}
where $|0\rangle=|s_z=\frac{1}{2}\rangle$ and $|1\rangle=|s_z=-\frac{1}{2}\rangle$.
Thus, the cell state is expressed by a qubit state $|w\rangle$ and an integer $i$ running from 0 to $m$ that indicates ancilla states. We will consider the decoherence of our cell later. We take into account a larger Hilbert space there. We assume all of these ancilla states have different energies ($E_0,E_1,\ldots,E_m$) and do not depend on qubit state $w$. $E_0$, especially, is assumed to be the lowest (i.e., ground state). We use this ancilla state as a sleeping state of the cell. For this to work well, the operation temperature must be sufficiently lower than the energy separation of ancilla bits. 

The whole system is a 1-, 2-, or 3-dimensional network of these kinds of cells. Each cell has a definite number of connections to neighboring cells. Let us define the interaction between two neighboring cells: cell a and cell b. The interaction is switched on if and only if the ancilla states of these cells are $k$ and $l$, respectively:
\begin{eqnarray}
\{|a\rangle=|w_a,k\rangle_a\} \wedge \{|b\rangle=|w_b,l\rangle_b\}\ \text{   for two bit operation},
\end{eqnarray}
where $k,l(\not=0,1,2,3,4,5)$ are specific numbers for pair a and b. We designed this such that $E_k(\text{cell a})=E_l(\text{cell b})$. This selective interaction can be realized, for example, by the real space transfers of electron wave functions due to the transition in ancilla bits. In other words, two electrons in two cells approach each other under this specific condition (Fig. \ref{fig:ancilla}). 

The interaction may be of any kind. However, the spin exchange is also thought to be useful in this scheme. Then, variable $w$ denotes the state of actual electron spin. The unitary operator, $U_{ab}(\theta)$, becomes
\begin{eqnarray}
U_{ab}(\theta):
\left\{
\begin{array}{lcl}
|0,k\rangle_a\otimes|0,l\rangle_b&\mapsto&|0,k\rangle_a\otimes|0,l\rangle_b\\
|0,k\rangle_a\otimes|1,l\rangle_b&\mapsto&\cos\theta|0,k\rangle_a\otimes|1,l\rangle_b-i\sin\theta|1,k\rangle_a\otimes|0,l\rangle_b\\
|1,k\rangle_a\otimes|0,l\rangle_b&\mapsto&\cos\theta|1,k\rangle_a\otimes|0,l\rangle_b-i\sin\theta|0,k\rangle_a\otimes|1,l\rangle_b\\
|1,k\rangle_a\otimes|1,l\rangle_b&\mapsto&|1,k\rangle_a\otimes|1,l\rangle_b.
\end{array}
\right.
\end{eqnarray}
The angle $\theta$ is proportional to the gating time in which both cells are in active states. The square root of swap ($\theta=\pi/4$) is useful in the construction of a controlled NOT gate \cite{Loss}.  
Since the swap gate has a single continuous parameter, it should be more powerful than a discrete controlled NOT gate. In other words, it is expected that circuit complexity is reduced. For the sake of the universality of quantum circuits, two-qubit gates for an arbitrary remote pair of qubits must be possible. However, in many physical implementations, this requirement should be extremely difficult. Although it is possible to arrange multiple two-bit operations for adjacent qubits so that they mimic single-operations for a pair of remote qubits, it costs a lot in terms of steps of gate. The swap operation means $U_{ab}(\pi/2)$ in a narrower sense. This operation is especially convenient since it completely exchanges the complex amplitudes of two qubits completely in a single step. Thus, the total number of steps necessary for some computational tasks should become much fewer. As easily understood, this interconnection problem is influenced by the average number of connections from a cell. To increase this value, higher dimensional structures are effective since the average distance between any qubit pair in a N-qubit circuit scales as $N^{1/d}$, where $d$ is the dimensionality of the network.

Next, we consider single-qubit operations. As in the standard scheme, a rotation with an arbitrary angle is facilitated. In our scheme, the rotation about the x-, y- and z-axes is switched on if and only if the ancilla state is 2, 3 and 4, respectively. Unitary operators are
\begin{eqnarray}
\begin{array}{lcl}
R_x(\theta):|w,2\rangle&\mapsto&|R_x(\theta)w,2\rangle\\
R_y(\theta):|w,3\rangle&\mapsto&|R_y(\theta)w,3\rangle\\
R_z(\theta):|w,4\rangle&\mapsto&|R_z(\theta)w,4\rangle .
\end{array}
\end{eqnarray}
For an arbitrary rotation, only two kinds of rotations with fixed axes, $R_x(\theta)$ and $R_y(\theta)$, are necessary since $R_z(\theta)$  can be synthesized by combining them. However, the full set of axes of rotations is effective for both the simplicity of circuit design and the reduction of circuit complexity.
A local magnetic field can rotate the spin. One possible method is described later.

The single-bit phase shift is a convenient operation for some quantum algorithms. The controlled phase shift, which is useful in Grover's algorithm, can be constructed simply by using the controlled NOT, rotation, and phase shift. The phase shift $\Phi(\phi)$ is switched on if and only if the ancilla state is 1. The unitary operator is
\begin{eqnarray}
\Phi(\phi):|w,1\rangle\mapsto e^{i\phi}|w,1\rangle.
\end{eqnarray}

The qubits must be measured at the end or during the course of calculations. We implement an ancilla bit for the measurement. A qubit is measured to be $|1\rangle$ if the ancilla state is 5 and $|0\rangle$ if the ancilla state is 0. The measurement process is described as
\begin{eqnarray}
c_0|0,0\rangle + c_1|1,5\rangle \to |0,0\rangle \ \text{or}\  |1,5\rangle.
\end{eqnarray}
This procedure is explained later in detail. Thus, in our scheme, only the sequence of $\pi$-pulses is necessary for operation, except in the final step of the measurement procedure. 

Now, the condition for the contraction of the Hilbert space introduced in eq. (\ref{contract}) is considered. The contraction is made in two steps. First, state vectors are restricted to
\begin{eqnarray}\label{contract2}
\begin{array}{lcl}
|w,\ 0,0\rangle&=&|w,0,0,\ldots,0\rangle\\
|w,z_1,1\rangle&=&|w,z,0,\ldots,0\rangle\\
|w,z_2,2\rangle&=&|w,0,z,\ldots,0\rangle\\
&\vdots&\\
|w,z_m,m\rangle&=&|w,0,0,\ldots,z\rangle,

\end{array}
\end{eqnarray}
where $z_1, z_2,\cdots,z_m \not = 0$. 
Namely, we ignore states like $|w,0,\ldots,0,z_k,0,\ldots,0,z_l,0,\ldots,0\rangle$. This means that the cell reduces to the product of a qubit and an $m+1$-level system such that the possibility of a large amplitude in more than two excited levels is neglected. As a simple model, a three-site, one-electron system is examined to find the necessary conditions for this assumption to hold. The hamiltonian is
\begin{eqnarray}
\hat{H}=E_ln_l+E_cn_c+E_rn_r+t(a_l^{\dagger}a_c+a_c^{\dagger}a_l-a_r^{\dagger}a_c-a_c^{\dagger}a_r)+s(a_l^{\dagger}a_r+a_r^{\dagger}a_l),
\end{eqnarray}
where suffixes l, c and r mean the left, center, and right dots. Figure \ref{fig:dotindep} shows the product of the probabilities of occupation in two different ancilla dots for one of the excited levels as functions of (a) direct transfer $s$ and (b) the detuning of site energies in both dots $|E_l-E_r|$. From these results, we are led to simple conditions: (1) direct transfer $s$ between different ancilla dots must be very small; (2) site energies of ancilla dots must have a sufficiently large difference (0.05 to 0.1 in this example). Both conditions determine the upper limit of the number of ancilla bits in a cell. It should be less than about 10. If the number of ancilla bits in a cell is small, the first condition can be fulfilled easily by the appropriate arrangement of ancilla dots since the tunneling probability decreases exponentially with the barrier thickness. 

The second contraction from eq. (\ref{contract2}) to eq. (\ref{contract}) is due to the operation scheme where ancilla bits are used not as qubits but as classical bits, and we always put the cell in either the completely active $|1\rangle$ or completely sleeping $|0\rangle$ state and not in a superposed state. Therefore, all $z_k$ take 0 or 1. For this condition to hold, $\pi$-pulse for the activation of ancilla must be accurate in frequency and duration. Furthermore, damping is utilized to "refresh" the ancilla state. This technique will be explained more clearly later.

The cell state is the tensor product of a qubit state and ancilla states. Both freedoms have their specific interaction with environments. In general, ancilla bits are far more strongly coupled to its environment than the qubit. We neglect the direct coupling of qubit to its environment and investigate indirect decoherence via a single ancilla bit, for simplicity. As the hamiltonian for qubit operation, we assume
\begin{eqnarray}
H_{op}=
\frac{J}{4}
\left[
\begin{array}{rr}
-1 & 2\\
2 & -1
\end{array}
\right]
\end{eqnarray}
in the Hilbert space ${\cal{H}}_q$ for a qubit. This is a submatrix of a swap hamiltonian. The creation operator $a^{\dagger}_c$, the annihilation operator $a_c$, and $\pi$-pulse operator $U_{\pi}$ are
\begin{eqnarray}
a^{\dagger}_c=
\left[
\begin{array}{rr}
0 & 1\\
0 & 0
\end{array}
\right],\ 
a_c=
\left[
\begin{array}{rr}
0 & 0\\
1 & 0
\end{array}
\right],\ \text{and}\ 
U_{\pi}=
\left[
\begin{array}{rr}
0 & 1\\
-1 & 0
\end{array}
\right]
\end{eqnarray}
in the Hilbert space ${\cal{H}}_a$ for the ancilla. The operation hamiltonian acting on the product space ${\cal{H}}_q\otimes{\cal{H}}_a$, denoted by tilde, is
\begin{eqnarray}
\tilde{H}_{op}=
\left[
\begin{array}{cc}
I & 0\\
0 & H_{op}
\end{array}
\right],
\end{eqnarray}
where I is the unit matrix. We examine the following operation procedure:
\begin{eqnarray}
\tilde{\rho}_i=|w,0\rangle\langle w,0| \stackrel{\tilde{U}_{\pi}}{\longrightarrow}|w,1\rangle\langle w,1|\stackrel{\tilde{U}_{op} \atop \text{decoherence}}{\longrightarrow}\tilde{\rho}\stackrel{\tilde{U}_{\pi}}{\longrightarrow}\tilde{\rho}^{\prime}\stackrel{\text{damping}}{\longrightarrow}\tilde{\rho}_f=\rho_f\otimes|0\rangle \langle0|_a,
\end{eqnarray}
where $\tilde{U}_{op}$ is a unitary evolution generated by $\tilde{H}_{op}$. We assume $\pi$-pulse is accurate. In the final step, only by leaving the cell alone for approximately same period as the operation, the ancilla bit resets to $|0\rangle$. This supports the contraction of the Hilbert space discussed earlier and keeps the ancilla bit as a classical bit. The second step in this procedure is described by a von Neumann equation with dissipation terms.
\begin{eqnarray}
\dot{\tilde{\rho}}=-\frac{i}{\hbar}[\tilde{H}_{op},\tilde{\rho}]-\frac{1}{2\tau_a} \{[a_c^{\dagger},a_c\tilde{\rho}]+h.c.\},
\end{eqnarray}
where $\tau_a$ is the coherence time for an ancilla bit. Figure \ref{fig:purity} shows the final purity for a full spin-flip operation as a function of the inverse time ratio $R_t^{-1}$ (gating time over coherence time), starting from $|0,0\rangle\langle 0,0|$ (the solid curve). The broken curve and dotted curve are final purities for a simple two-level system starting from $|1\rangle$ and $(|0\rangle+|1\rangle)/\sqrt{2}$, respectively. 
Thus, the indirect spin decoherence rate is in the same order as the ancilla decoherence rate. However, this decoherence occurs only in a gated period. If the duty ratio $R_d$ of active time to total time is much smaller than unity, the effective decoherence time $\tau_a/R_d$ becomes very long.
In quantum circuits, only single- and two-qubit gates are used. If parallelism is not used, in a N-qubit circuit, the duty ratio ranges from 1/N to 2/N. Thus, the larger the circuit, the more efficient this scheme becomes. The disadvantage of excess decoherence due to the incorporation of ancilla bits should be largely compensated. 

Our scheme is not restricted to a particular physical system. It may be possible in molecular quantum computers or trapped-ion systems. However, we focus on the more realistic quantum-dot based implementation. In a quantum dot system, a qubit is equipped with an ancilla bit only by locating another dot in the proximity of the main dot. Therefore, our cell is made of m+1 coupled dots. To have the lowest energy level, the main dot has the largest size. Instead, material with the lowest conduction band edge may be used. The main dot is surrounded by ancilla dots of different sizes. Resonant optical $\pi$-pulses transfer electrons from the main dot to a specific ancilla dot, and vice versa.

 Figure \ref{fig:device} shows an example of the proposed scheme with a rather economical cell construction strategy. In this example, only seven ancilla dots are used in each cell. It adopts the design of a 2-dimensional square lattice. The $R_x$- and $R_y$-dots denoted by 2 and 3 are located along the positive and negative z-axes, respectively. The measurement dot denoted by 5 is above the $R_x$-dot. The $R_z$- and $\Phi$-dots are omitted. Ancilla dots for single-bit operations are away from each other and neighboring cells. On the contrary, the four ancilla dots denoted by 6, 7, 8, and 9 for two-bit operations are located toward the neighboring cells, like chemical bonds in molecules. The distances between ancilla dots responsible for this operation (e.g., 8th dot in the (1,0)-cell and 6th dot in the (0,0)-cell) are reduced for kinetic exchange interaction to work. Furthermore, the matrix element for the dipole moment, $M_8^{(1,0)}$, between ground state $|w,0\rangle_{(1,0)}$ and active states $|w,8\rangle_{(1,0)}$, and $M_6^{(0,0)}$, between $|w^{\prime},0\rangle_{(0,0)}$ and active states $|w^{\prime},6\rangle_{(0,0)}$, is designed so that $|M_8^{(1,0)}|=|M_6^{(0,0)}|$ in order to synchronize the activation. The matrix element for the dipole moment can be tuned by designing the sizes of dots and the distance between them since 
$M_i\equiv\langle w,i|e \bbox{r}|w,0\rangle=\int d^3\bbox{r} \phi_i(\bbox{r})e \bbox{r}\phi_0(\bbox{r})$, where $\phi_0$ and $\phi_i$ are the wavefunctions for electrons confined in the main and i-th ancilla dots. 

As the energy levels differ, ancilla bits can be distinguished by their frequency of optical $\pi$-pulse. As the extension of this multiplicity in a wavelength domain, the different dots have different energies whether they are in the same cell or different cells. However, the differences of eigenenergy in different cells may be much smaller than those in the same cell since small detuning is sufficient for suppressing Rabi oscillations. This fact will help to save frequency resources. Furthermore, dots in remote cells that have separations larger than the radius of the laser spot may have identical spectrums. 

A local magnetic field for $R_{x,y,z}$-dot is realized, of course, simply by placing the nanometer-sized permanent magnet next to the dot. However, since it is hard to shield the magnetic field completely, spins outside of $R_{x,y,z}$-dots can also be influenced by it. To solve this problem, we propose to adopt special configuration of ferromagnets embedded in our network as shown in Fig. \ref{fig:layout}. We use $R_{\xi,\eta}$ instead of $R_{x,y}$, where $x=(\xi-\eta)/\sqrt{2}, y=(\xi+\eta)/\sqrt{2}$. Small ferromagnets lie laterally such that they connect the $R_{\xi,\eta}$-dots of diagonally neighboring cells. Thus, local magnetic fields are directed in the positive $\xi$-direction for the upper $R_{\xi}$-dot layer and positive $\eta$-direction for the lower $R_{\eta}$-dot layer. The magnetization of small ferromagnets is done by applying an external static magnetic field toward the positive y-direction. The leakage of magnetic flux can be made small by reducing the gap between poles. This device structure is not hard to fabricate since we have no semiconductor quantum dots on the top of the ferromagnetic material.
We also propose to use high g-factor material for the $R_{\xi,\eta}$-dots for reducing the effect of leakage of the magnetic field. In a relatively weak magnetic field, the electron spin in a $R_{x,y,z}$-dot can make rapid precession without affecting the spin in other dots with a small g-factor. Diluted magnetic semiconductors such as $\text{Ga}_x\text{Mn}_{1-x}\text{As}$ are the interesting candidates for this purpose, since it can be grown epitaxially on GaAs. The evidence of strong p-d exchange interactions have been suggested theoretically and experimentally. This may support the usage of holes instead of electrons in our cells.

As for the measurement of the qubit, the help of $R_{\xi}$-dot is needed along with a measurement dot. The measurement dot must have lower energy level than the $R_{\xi}$-dot. The measurement procedure is as follows; 
\begin{eqnarray}
|\text{initial}\rangle = c_1|\uparrow,0\rangle+c_2|\downarrow,0\rangle\\[3mm]
\stackrel{\pi_1}{\longrightarrow}c_1|\uparrow,2\rangle+c_2|\downarrow,0\rangle\label{eq:pi1}\\[3mm]
\stackrel{\pi_2}{\longrightarrow}c_1|\uparrow,5\rangle+c_2|\downarrow,0\rangle\label{eq:pi2}\\[3mm]
\stackrel{\text{measurement}}{\longrightarrow} |\uparrow,5\rangle \ \text{ or }\  |\downarrow,0\rangle.\label{eq:measure}
\end{eqnarray}
First, using a $\pi$-pulse with a line spectrum tuned to the upspin level in the $R_{\xi}$-dot, the corresponding component of electron wavefunction is transferred from the main dot to the $R_{\xi}$-dot: eq. (\ref{eq:pi1}). Next, using second $\pi$-pulse, the same component of electron wavefunction is transferred from the $R_{\xi}$-dot to the measurement dot: eq. (\ref{eq:pi2}). Finally, the polarization of the cell is detected by capacitively coupling the central island of a single electron transistor to the measurement dot: eq. (\ref{eq:measure}). The state in eq. (\ref{eq:pi2}) is very stable in the sense that the partition of probability among two dots (0 and 5) is not changed by the coulomb interactions of many electrons flowing through tunnel junctions of the single electron transistor since direct transfers between these dots is forbidden. Although the phase memory would be lost very quickly in the course of this last step, amplitude memory is sustained until we finally observe the output voltage of the single electron transistor.

In conclusion, the quantum network structure with ancilla bits in every cell was proposed. The whole circuit works only with the help of external optical pulse sequences. External switchings of electric and magnetic field are not necessary. In operation, an ancilla bit is activated and autonomous single- and two-bit operations are made. In the sleep mode of cell, the decoherence of a qubit is negligibly small. In the active state, indirect decoherence due to the coupling of an ancilla bit to its environment is in the same order of magnitude as simple ancilla decoherence. However, the fact that the duty ratio of each bit scales as $N^{-1}$ improves effective decoherence time. A device structure using a quantum dot array with possible operation and measurement schemes was also proposed.

\begin{figure}
 \caption{The selective interaction between cell a and cell b by the real space transfers of electrons into k-th and l-th ancillas due to an optical $\pi$ pulse.}
 \label{fig:ancilla}
 \end{figure}

\begin{figure}
 \caption{Product of probabilities of occupation in both dots as functions of (a) direct transfer $s$ and (b) energy difference of levels $|E_l-E_r|$ in both dots.}
 \label{fig:dotindep}
 \end{figure}

\begin{figure}
 \caption{Final purity $s$ with full spin flip operation as function of inverse time ratio $R_t^{-1}$ (gating time over coherence time) starting from $|0,0\rangle\langle 0,0|$ (the solid curve). The dotted curve and broken curve are purities for a simple two-level system starting from $|1\rangle$ and $|0\rangle+|1\rangle$, respectively.}
 \label{fig:purity}
 \end{figure}

 \begin{figure}
 \caption{Example of 2-dimensional square lattice version of proposed scheme. Seven ancilla dots are used in each cell. The $R_x$- and $R_y$-dots denoted by 2 and 3 are located along the positive and negative z-axes. The measurement dot denoted by 5 is above the $R_x$-dot. The four ancilla dots for the two-bit operation denoted by 6, 7, 8, and 9 are located towards the neighboring cell. }
 \label{fig:device}
 \end{figure}

 \begin{figure}
 \caption{Proposed configuration of nanoscale ferromagnets embedded in network of Fig. 3. The larger circles are the central columns of cells. The rectangles with the solid (broken) line indicate ferromagnets in the upper (lower) layer of $R_{\xi}$-dots ($R_{\eta}$-dots). The ferromagnets are magnetized by an external magnetic field toward positive y direction.}
 \label{fig:layout}
 \end{figure}
 
\begin{center}

\begin{figure}[htbp]
\input epsf
\epsfxsize=150mm \epsfbox[30 209 528 567]{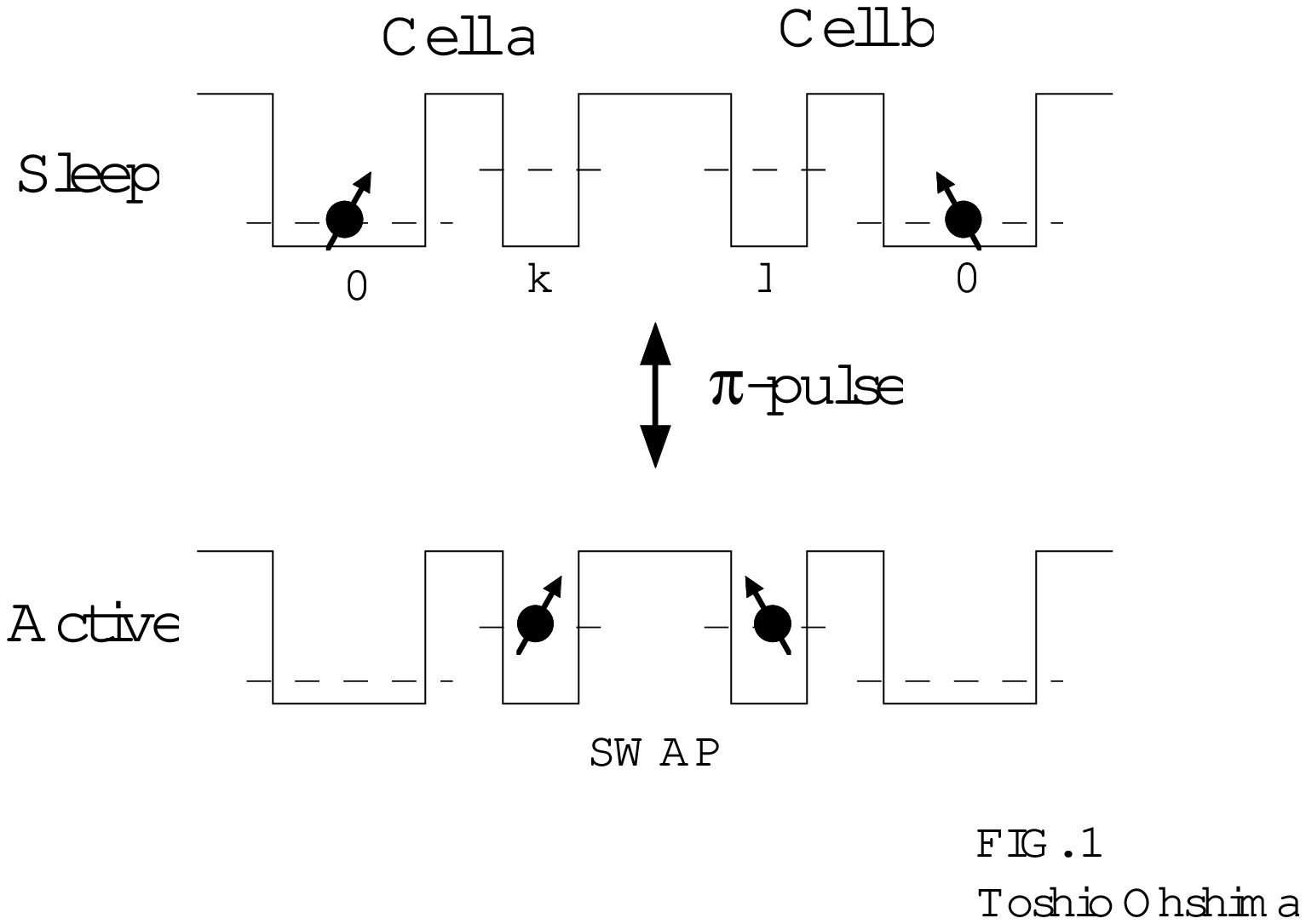}
\end{figure}

\begin{figure}[htbp]
\input epsf
\epsfxsize=150mm \epsfbox[6 6 591 838]{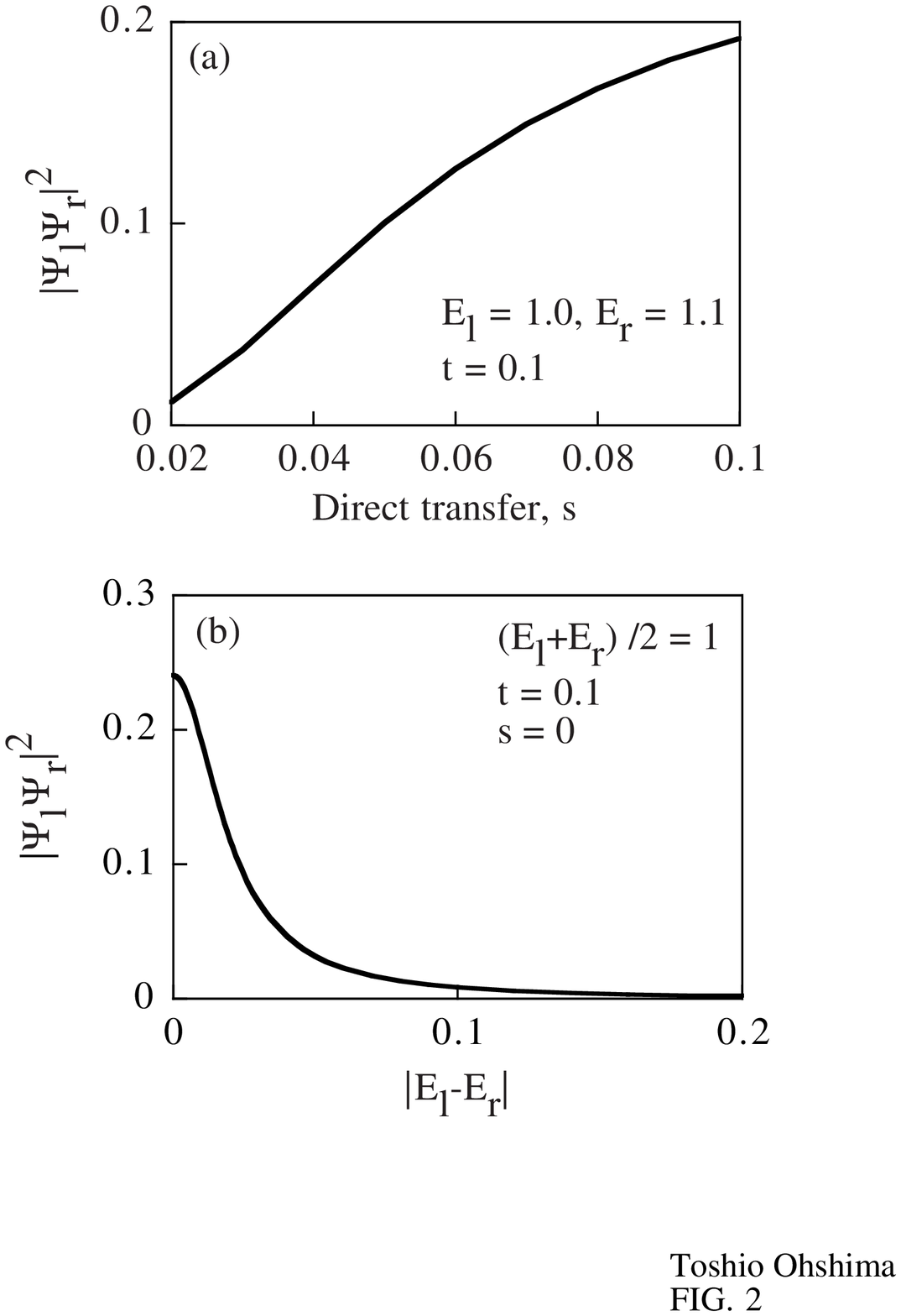}
\end{figure}

\begin{figure}[htbp]
\input epsf
\epsfxsize=150mm \epsfbox[30 31 567 811]{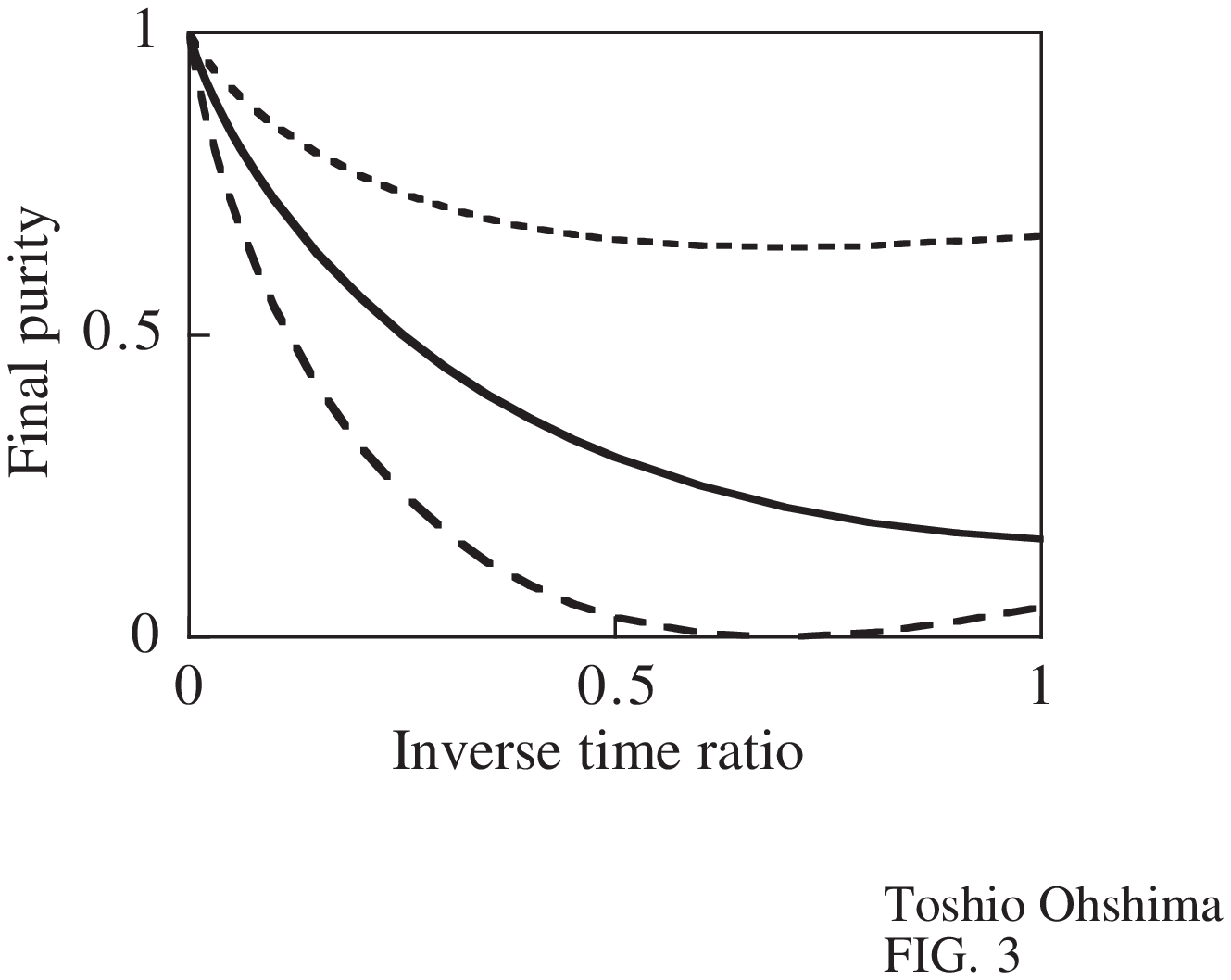}
\end{figure}

\begin{figure}[htbp]
\input epsf
\epsfxsize=150mm \epsfbox[104 35 790 501]{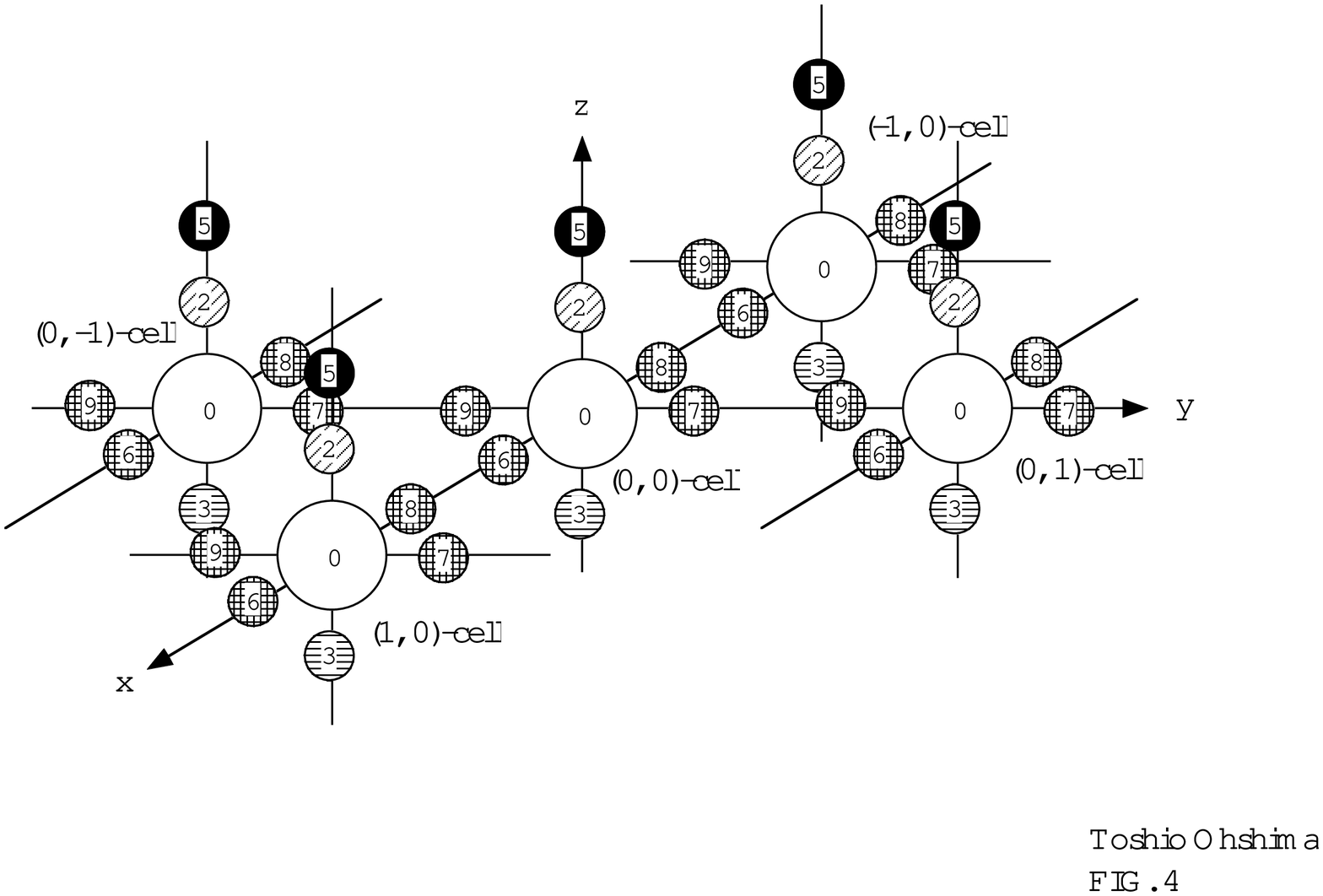}
\end{figure}

\begin{figure}[htbp]
\input epsf
\epsfxsize=150mm \epsfbox[3 29 809 591]{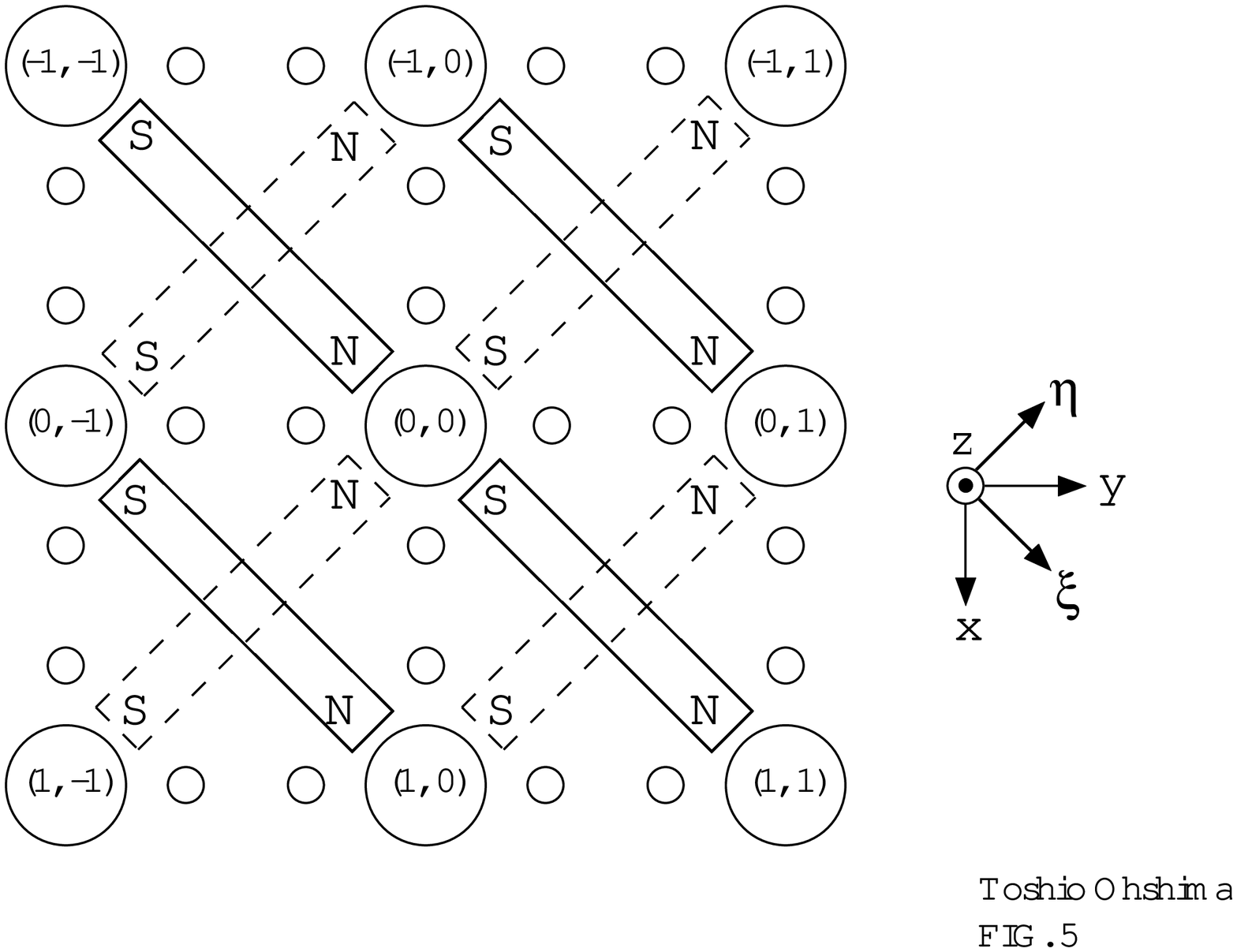}
\end{figure}

\end{center}

\end{document}